\documentstyle[preprint,aps]{revtex}
\begin{document}
\draft
\title{Linear response and collective oscillations in superconductors with
$d$--wave pairing}
\author{S.N. Artemenko, A.G. Kobelkov}
\address{Institute for Radioengineering and Electronics of
the Russian Academy of
Sciences, 103907 Moscow, Russia}

\date{\today}
\maketitle
\begin{abstract}
Simple equations for the linear response of layered superconductors with
$d$--wave symmetry of the order parameter are derived by means of kinetic
equations for Green's functions. Responses to solenoidal and potential
electric fields have different frequency dependencies. The damping of plasma
oscillations of superconducting electrons is determined by dielectric
relaxation and is small. Relaxation of branch imbalance determined by
elastic scattering is large enough to make the Carlson--Goldman mode in
$d$--wave superconductors overdamped.
\end{abstract}
\pacs{74.25.Fy, 74.25.N1, 74.20.Mn}

Many evidences for $d$--wave symmetry (or near $d$--wave symmetry) of the
superconducting order parameter in layered high--T$_c$ superconductors were
given last years by measurements of the Josephson effect
\cite{Woll93,Igu94,Br96}, by microwave experiments \cite{Har93}, by
high-resolution angle-resolved spectroscopy \cite{Yok} and by other methods.
On the other hand, theoretical studies confirm a compatibility of many
experimental data with $d$--wave symmetry of the superconducting gap in
high--T$_c$ superconductors (see \cite{Sig92,Hir} and references therein).
Many properties of the $d$-wave superconductors are expected to be different
>from those of conventional superconductors, especially the effects related to
quasiparticles and their relaxation, since due to the nodes of the $d$--wave
order parameter the quasiparticle density is never exponentially small. Here,
we study theoretically the effects related to electric field, and collective
oscillations in $d$--wave superconductors.

Typically, calculations of the linear response of the superconductors assume
the response to a solenoidal (transverse) electric field which can be
expressed in terms of time derivative of the vector potential, paying less
attention to a potential (longitudinal) electric field which is determined as
the gradient of the scalar potential. However, the linear response of
superconductors depends on the origin of the electric field.  Electric field
created by variations of the current density and magnetic field in time, and
related to the Faraday's law, creates in a superconductor only the
perturbations of the electronic distribution with the antisymmetric angle
dependence. Such a perturbation is limited by the momentum relaxation, like
in a normal metal. On the other hand, the potential electric field which is
related to perturbations of the charge density and to the Coulomb's law
creates, in addition, the branch imbalance \cite{TC,T,AV}, {\em i.  e.} the
difference between the densities of electron--like and hole--like
quasiparticles. Thus, relaxation of the branch imbalance is involved in the
linear response of a superconductor to the potential electric field as well.
The branch imbalance is characterized by the gauge invariant scalar potential
$\mu =(1/2) (\partial \chi /\partial t)+\Phi$ where $\chi$ is the phase of
the order parameter and $\Phi$ is the electric potential. Potential $\mu$ can
be interpreted as the nonequilibrium shift of the chemical potential of the
normal carreers. Furthermore, the superconducting momentum, ${\bf P}_s =
(1/2)\nabla \chi - {\bf A}$ where ${\bf A}$ is the vector potential, plays a
role of the gauge invariant vector potential (see \cite{AV} and references
therein). Then the electric field is expressed in terms of the gauge
invariant potentials as
\begin{equation}
{\bf E} = - \nabla \mu+ \frac{\partial {\bf P}_s}{\partial t}. \label{E}
\end{equation}
So, two contributions to the electric field in (\ref{E}) produce different
kinds of the perturbations of the electronic distribution in a
superconductor. We shall consider the first term of this expression as the
potential part of the electric field. The second term related to the
variations of the superconducting current in time we shall call the
solenoidal part of the electric field. Strictly speaking, the second term in
(\ref{E}) does not satisfy the usual definition of the solenoidal field
because its divergence does not vanish exactly in all the cases.
Nevertheless, we call it solenoidal or transverse field because it is needed
to describe the purely solenoidal field, and it is related to the response to
the electric field created by the time dependent magnetic field.

The potential electric field must be taken into account in the problem of the
linear response since it is important in collective oscillations and it
appears in nonuniform and anisotropic systems, even if externally applied
field is purely solenoidal. In order to study the problem we calculate a
linear response of $d$--wave superconductors to the gauge invariant vector
and scalar electromagnetic potentials, and derive simple and physically
transparent expressions for the charge and current densities from the
equations for quasiclassical Green's functions using the nonequilibrium
approach by Keldysh \cite{K}. Such an approach enables us to take into
account both momentum and branch imbalance relaxation rates. The latter
enters the generalized conductivity describing the response to the potential
electric field. Then we use these expressions to study plasma oscillations of
superconducting electrons and the Carlson--Goldman mode in $d$--wave
superconductors.

To calculate the linear response of layered superconductors with $d$--wave
pairing, we start with the equations for Green's functions in Keldysh
technique, using the slightly modified approach by Larkin and Ovchinnikov
\cite{LO}. We use two different ways to describe layered superconductors. In
the first approach we use a continuous representation, considering an
anisotropic metal with the band motion of electrons in the direction
perpendicular to the layers such that $t_\perp \gg\nu$, where $t_\perp$ is
the overlap integral describing the electron spectrum in perpendicular
direction, $\epsilon_\perp =2t_\perp\cos{dp_\perp}$. Here $d$ is the lattice
constant in the perpendicular direction, and $\nu$ the momentum scattering
rate along the layers. In the second approach we use similar equations
\cite{A80,AK} for layered superconductors in the discrete Wannier
representation, considering the hopping conductivity regime between the
layers, {\em i. e.} $t_\perp \ll\nu$, which corresponds to the case of
Josephson interlayer coupling. The second approach bears some similarity to
the interlayer diffusion model \cite{Gr93} in which the interlayer coupling
is mediated through incoherent hopping processes with $t_\perp$ neglected. In
both cases we assume a $d$--wave superconducting order parameter: thus we do
not address the question of the microscopic nature of the interaction
resulting in such a symmetry.

To derive the equations in the continuous representation we subtract, similar
to \cite{LO}, from the equation for matrix Green's function in Keldysh
representation its conjugated equation. Then we integrate the resulting
equation over $\xi=p_\|^2/2m - \varepsilon_F$, where ${\bf p}_\|$ is the
component of the momentum parallel to the layers. Thus we obtain equations for
the retarded (advanced) Green's functions, $g^{R(A)}$, and for $g^K$, which is
related to the electron distribution function. Each of these functions is a
matrix in Nambu space and depends on coordinates, energies, perpendicular
component $p_\perp$ of the momentum, and on the angle $\phi$ of ${\bf p}_\|$.

In the linear approximation with respect to the external perturbation, the
equation for the anomalous Green's function, $g^{(a)}$, defined by $g^K =
g^R(\varepsilon, \varepsilon') \tanh{(\varepsilon'/2T)} -g^A(\varepsilon,
\varepsilon') \tanh{(\varepsilon/2T)} +g^{(a)}(\varepsilon, \varepsilon')$ has
the form
\begin{eqnarray}
{\bf v} \nabla g^{(a)} -[\varepsilon_+\sigma_z +
\Delta(\phi) i\sigma_y]g^{(a)} +g^{(a)}[\varepsilon_-\sigma_z +
\Delta(\phi) i\sigma_y] - \nonumber \\
( \Sigma^R g^{(a)} - g^R\Sigma^{(a)} + \Sigma^{(a)} g^A - g^{(a)}\Sigma^A )=
\alpha[({\bf v P}_s \sigma_z + \mu)g^A - g^R({\bf v P}_s \sigma_z + \mu)].
\label{ga}
\end{eqnarray}
Here ${\bf v}$ is the electron velocity at the Fermi surface, $\Delta(\phi)$
is the amplitude of the order parameter, $\alpha =\tanh{\varepsilon_+/2T}-
\tanh{\varepsilon_-/2T}$, $\sigma_{y,z}$ are Pauli matrices, and the
unperturbed retarded and advanced Green's functions,
$g^{R(A)}$, depend on shifted energies $\varepsilon_+ =\varepsilon+\omega/2$
and $\varepsilon_-=\varepsilon-\omega/2$, respectively. The self--energy
parts, $\Sigma^\iota$, are given by
\begin{equation}
\hat{\Sigma}^\iota = \int^{\pi/d}_{-\pi/d}\frac{dp_\perp'}{2\pi
/d}\int_{0}^{2\pi}
\frac{d\phi'}{2\pi} \nu (p_\perp,\phi;p_\perp',\phi')
\hat{g}^\iota(p_\perp',\phi'), \label{I}
\end{equation}
where $\iota=$R, A, or K; $\nu$ is the elastic scattering rate in the normal
state. Strictly speaking, (\ref{I}) describes the impurity scattering in Born
approximation, but it can be applied also to elastic scattering by phonons,
since the related self--energy part acquires the form (\ref{I}) when one
neglects phonon frequencies in comparison with electron energies in
delta--functions describing the energy conservation law in scattering
processes. Using Born approximation we neglect low--energy quasiparticle
bound states created by impurities (see \cite{Gr96} and references therein),
and, hence, our results are applicable provided typical energies of
quasiparticles are larger than the bandwidth of the impurity induced bound
states, $T > \sqrt{\Delta \nu}$.

Now we consider the momentum dependence of the scattering rate $\nu$. As it
will be seen below, the in--plane scattering results in pair--breaking similar
to magnetic impurities in s--wave superconductors, while the interlayer
scattering does not affect the gap. In addition, from the conductivity
anisotropy data in high--T$_c$ superconductors one may expect, that
corresponding components of $\nu$ have different temperature dependencies.
Having this in mind we consider a simple model for momentum dependence of
$\nu$, which takes into account different scattering rates in different
directions:
\begin{equation}
\nu (p_\perp,\phi;p_\perp',\phi') = \nu_i
+\nu_\perp\delta(\phi-\phi').\label{nu}
\end{equation}
Here $\nu_i$ describes the isotropic scattering, and $\nu_\perp$ is related
to the scattering in the perpendicular, interlayer, direction.

Using (\ref{I}) and (\ref{nu}) we obtain for the unperturbed retarded
(advanced) Green's functions in (\ref{ga}) the implicit relations
$g^{R(A)}=\sigma_z a^{R(A)} +i\sigma_y b^{R(A)}$, where $a^{R(A)}=(\varepsilon
+ i\nu_i\langle a^{R(A)}\rangle_\phi/2)/\xi^{R(A)}$, $b^{R(A)}=\Delta
(\phi)/\xi^{R(A)}$. The brackets $\langle ...\rangle$ mean averaging over
variables mentioned in the subscript, and $\xi^{R(A)}=\pm \sqrt{(\varepsilon +
i\nu_i\langle a^{R(A)}\rangle_\phi/2)^2 -\Delta (\phi)^2}$. The equations for
perturbations of $g^{R(A)}$ can be obtained from (\ref{ga}) by replacing
$\alpha$ by 1, changing all the superscripts for $R(A)$, and omitting the two
last terms in the collision integral.

The main differences in the equations for the Green's functions in the
discrete representation (see \cite{AK}) are the following. Green's functions
become matrices in layer indices, and the potentials depend on the layer index
as well. The first term in the l.h.s. of (\ref{ga}) in the discrete
representation is to be replaced by $t_\perp \sum_{i=\pm 1}(A_{nn+i}
g^{(a)}_{n+im}- g^{(a)}_{nm+i}A_{m+im})$, describing the interlayer
interaction, with $ A_{nm}=\cos{ ( \chi_n-\chi_m)/2 }+ i\sigma_z
\sin{(\chi_n-\chi_m)/2}$. Averaging in the collision integral is performed
over the angle $\phi$ only.

We solve the linearized equations for Green's functions for the case of
smoothly varying perturbations $|{\bf qv}| \ll \nu_i$, {\em i. e.} when
changes of all variables along the distance of the order of the mean free
path along the layers are small. This case covers the most interesting range
of frequencies, because characteristic values of $1/q$ are determined either
by the magnetic penetration lengths (at low frequencies) which are typically
larger, than the mean free path in high-T$_c$ superconductors, or (at high
frequencies) by the skin--effect length, which is also large provided the
frequency is below the range of the anomalous skin-effect.

Perturbations of charge density are determined by ${\rm Tr} \langle
g^{(a)}\rangle$ integrated over energies. Quasiparticle current densities in
the directions parallel and perpendicular to the conducting layers are
proportional to integrals of ${\rm Tr}\sigma_z\langle {\bf v} g^{(a)}\rangle$
and of ${\rm Tr}\sigma_z\langle v_z g^{(a)}\rangle$, respectively.
Superconducting currents are determined by similar terms with retarded and
andvanced Green's functions in the relation between $g^K$ and $g^{(a)}$. We
calculate current and charge densities assuming the clean limit, $T_c \gg
\nu_i$, since in the opposite dirty limit a superconductor is in a gapless
state. In the case of frequencies $\omega$ much smaller than
the amplitude of the gap, $\Delta$ the linear response can be presented in
a simple and physically transparent form:
\begin{eqnarray}
-i\omega \rho =-i\omega\gamma \frac{\kappa^2}{4\pi} \mu+ (\sigma_{2l}q^2 +
\sigma_{2t}k^2)\mu + \omega(\sigma_{1l}{\bf qP}_l +\sigma_{1t}kP_t),
\label{mu}\\
{\bf j}_l =\frac{c^2}{4\pi \lambda_l^2}{\bf P}_l- i(\omega\sigma_{0l}{\bf
P}_l +\sigma_{1l}{\bf q}\mu), \label{jl}\\
j_t =\frac{c^2}{4\pi \lambda_t^2} P_t- i(\omega\sigma_{0t} P_t +
\sigma_{1t} k\mu), \label{jt}
\end{eqnarray}
where ${\bf P}_l$ and $P_t$ are parallel and perpendicular to the layers
components of ${\bf P}_s$, ${\bf q}$ and $k$ are parallel and perpendicular
components of wave vector, $\kappa^{-1}$ is the Thomas--Fermi screening
radius, and $\lambda_{l(t)}$ are the penetration lengths for a
superconducting current parallel (perpendicular) to the layers.

The first terms in (\ref{jl},\ref{jt}) describe the supercurrents, while the
last terms are related to the quasiparticles. Taking into account that the
electric field expressed in terms of the gauge invariant potentials is given
by (\ref{E}) we see, that the simple expression ${\bf j}=\hat{\sigma} {\bf
E}$ for quasiparticle contributions to the currents is not valid: generalized
conductivities $\sigma_{n\alpha}$ are different for the contributions from
scalar and vector potentials to the electric field. This implies different
responses to the potential and to the solenoidal electric fields. Then note
that according to (\ref{mu}) the changes of the charge density are determined
by time variations of the potential $\mu$, which is related to the
electrone-hole imbalance (cf. \cite{T,AV}), and by space variations of the
quasiparticle currents. Equation (\ref{mu}) plays a role of the continuity
equation for normal carreers.

The factor $\gamma$ and the generalized conductivities $\sigma_{n\alpha}$
($n=0,1,2$
and $\alpha=l,t$) depend on frequency:
\begin{eqnarray}
\gamma=1+ \int_{-\infty}^{\infty} d\varepsilon \frac{\omega
\langle a_0\rangle_\phi}{(\omega+i\nu_b)}\frac{dn_F}{d\varepsilon},
\label{gam}\\
\sigma_{n\alpha}=-\sigma_{N\alpha}
\frac{1}{\tau_\alpha}\int_{-\infty}^{\infty} d\varepsilon
\langle \frac{i\theta(|\varepsilon|-|\Delta(\phi)|) a_0^{1-2n}\omega^n}
{(\omega+ i\tilde{\nu}_\alpha) (\omega+i\nu_b)^n}
\frac{dn_F}{d\varepsilon}\rangle_\phi \label{s}
\end{eqnarray}
Here $\sigma_{N\alpha}$ is the normal state conductivity in direction $\alpha$,
$n_F$ is Fermi distribution function, and
$a_0=\varepsilon/\sqrt{\varepsilon^2-\Delta^2}$; $\tau_l\equiv
1/\nu_i=1/\nu_l$ and $\tau_t=1/(\nu_i+\nu_\perp)$ are the momentum scattering
times of electrons in the normal state for longitudinal and transverse
directions, and $\tilde{\nu}_l=\nu_l\langle a_0\rangle_\phi$ and
$\tilde{\nu}_t=\tilde{\nu}_l + \nu_\perp/a_0$ are energy dependent effective
scattering rates of quasiparticles for corresponding directions. Finally,
$\nu_b= \nu_l\langle \Delta^2(\phi) a_0/ \varepsilon^2\rangle_\phi$ is the
effective branch imbalance relaxation rate. It is well-known that in s-wave
superconductors the branch imbalance relaxes via inelastic scattering,
spin-flip scattering or due to anisotropy of the order parameter (for a
review see \cite{AV}). In the case of $d$-wave pairing the elastic scattering
is a main source of the branch imbalance relaxation.

In the studies of the linear response to a solenoidal field, the conductivity
$\sigma_{0l}$ in (\ref{jl}) is usually calculated. Our result for
$\sigma_{0l}$ agrees with the Born limit of the general expressions for the
conductivity obtained in \cite{Hir}.

Note that the conductivities in the transverse direction are determined by
contributions both from intralayer scattering and by interlayer scattering, so
that the effective scattering rate for the conductivity in transverse
direction is larger than the effective scattering rate for the in-plane
quasiparticle current.

Solution of the discrete equations in the limit of small phase differences
between the neighbouring layers gives results similar to (\ref{mu}-\ref{s});
it can be obtained from (\ref{mu}-\ref{jt}) substituting $P_t$ for $(\chi_n
-\chi_{n-1})/d$, and $v_z$ for $2t_\perp d$ with $\nu_\perp = 0$ and $\nu_i
\equiv\nu$ in (\ref{gam}-\ref{s}), so that conductivities in both directions
are determined by the same scattering rate.

Now we discuss the limits of low ($\Delta \gg T$) and high ($\Delta \ll T$)
temperatures.

1. $\Delta \gg T$: an important distinction from s-wave superconductors is
that the conductivities in (\ref{s}) are not exponentially small.

Consider, first, the linear response to the electromagnetic wave. For
simplicity, in order to get explicit expressions we consider the simplest
angular dependence of the gap parameter with the $d$-wave symmetry
$\Delta = \Delta_0 \cos{2\phi}$ .
At low temperatures, $\langle a_0\rangle_{\phi}=\varepsilon/\Delta_0$,
and the characteristic times for quasiparticles averaged over energies are
$\tilde{\tau}_l= \tau_l(\Delta_0/2T) \approx \tau_b \gg \tau_l$. The relative
density of superconducting component in $d$-wave superconductor is $N_s=1 -
(T/\Delta_0)\ln{4}= (\lambda(0)/\lambda(T))^2 \approx 1$.The factor $\gamma$
is also very close to 1, $\gamma = 1+2i\omega\tau_l$ for $\omega \ll 1/\tau_b$,
and $\gamma = 1- (T/\Delta_0)\ln{4}$ for $\omega \gg 1/\tau_b$. This leads to a
smaller contribution of scalar potential $\mu$ (though not exponentially small
as in s-wave superconductors), and we may omit the diffusion contribution to
the quasiparticle current densities. Then the current densities can be
presented as
\begin{eqnarray}
{\bf j_\alpha} =\frac{c^2}{4\pi \lambda_\alpha^2}{\bf P}_\alpha-
i\omega\sigma_{N\alpha}{\bf P}_\alpha R_{\alpha}(\omega) \label{jlow}\\
R_l=\int_{0}^{\infty} \frac{xdx}{(x-i\omega\tilde{\tau_l})\cosh^2{x}},
\label{Rl}\\
R_t=\frac{2\tau_l}{\pi \tau_t} \int_{0}^{\infty}dx \int_0^1
\frac{xdy}{(x-i\omega\tilde{\tau_l}+\nu_\perp\tilde{\tau_l}\sqrt{1-y^2})\cosh^2
{x}}, \label{Rt}
\end{eqnarray}

According to (\ref{jlow}-\ref{Rl}), at $\omega\tilde{\tau_l} \ll 1$, {\em i. 
e.}
when scattering is important, the decrease of the normal carreer density is
compensated by the decrease of the scattering rate of quasiparticles in
comparison with the normal state by the same factor $\propto T/\Delta$.
Furthermore, $R_l=1$, and the quasiparticle conductivity along the layers
even at low temperatures is the same as it would be in the normal state at
this temperature.  At higher frequencies ($\omega\tilde{\tau_l} \gg 1$)
$R_l=i(1-N_s)/(\omega\tau_l)$, scattering is not important and the current
density corresponds to a free motion of all electrons.

Consider now the conductivity in the transverse direction. Note that due to
the depairing action of the in-plane elastic scattering, these processes
contribute to the conductivity in the transverse direction. If $\nu_\perp \ll
1/\tilde{\tau_l}$ or $\nu_\perp \ll \omega$, we find
$R_t=R_l(\tau_l/\tau_t)$. Thus in this case the transverse conductivity is
determined by the in-plane scattering. At $\nu_\perp \gg 1/\tilde{\tau_l}$
and $\nu_\perp \gg \omega$, we obtain $R_t \propto(T/\Delta_0)
\ln{\nu_\perp^2/(\omega^2 +1/ \tilde{\tau_l}^2)}$
is small and can be neglected.

Using equations (\ref{jlow}-\ref{Rt}) one can easily calculate the surface
impedance of a $d$-wave superconductor. For a surface parallel to the
layers we obtain
\begin{equation}
\zeta=\frac{\omega \lambda_l}{c} \sqrt{\frac{1-(\omega/\omega_0)^2 -iR_t
\omega\tau_t +(q\lambda_t)^2}{(1-(\omega/\omega_0)^2 -iR_t \omega\tau_t)
(1-iR_l \omega\tau_l)}}, \label{z}
\end{equation}
where $\omega_0 = c/\lambda_t$ is the frequency of plasma oscillations for
an electric field perpendicular to the layers.

We conclude that, in spite of the large quasiparticle conductivity equal to
the normal-state conductivity, the damping terms at low temperatures are
always small, because the scattering of quasiparticles is important only at
$\omega < 1/\tilde{\tau_l} \ll 1/\tau_l$.

Now we discuss the spectrum of free oscillations, which can be calculated
inserting (\ref{mu}-\ref{jt}) into the Maxwell equations. The spectrum
of the weakly damped plasma mode in the long-wavelength limit is given by
an expression similar to that of the case of $s$-pairing \cite{AK}:
\begin{equation}
\omega^2= \omega_0^2
\left(\frac{1+k^2\lambda_l^2+q^2\lambda_t^2}{1+k^2\lambda_l^2} -
iR_t\omega\tau_t \right) \label{pl}
\end{equation}
The last term in (\ref{pl}) describes damping. For small frequencies limit it 
is
determined by the dielectric relaxation frequency $4 \pi\sigma_N$, which is
rather large. Nevertheless, the plasma oscillations survive, because the
damping is determined by $\sigma_N$ only at frequencies $\omega < T/(\tau
\Delta) \ll 1/\tau$, and becomes small at $\omega \approx \omega_0 >
T/(\tau \Delta) $.

2. $\Delta \ll T$: at high temperatures (but outside the gapless regime
$\Delta < \nu_l$) the branch imbalance relaxation rate is much smaller than
the elastic collision rate, and the conductivities (\ref{s}) depend on the
relation between frequency $\omega$ and $\nu_l(\Delta/T)^2$.

In the frequency range $\omega \gg \nu_l(\Delta/T)^2$ in $s$-wave
superconductors, where the factor $\gamma$ is real (see \cite{AV}), the
weakly damped Carlson-Goldman mode appears. In $d$-wave superconductors the
factor $\gamma = (\pi\Delta_0/2T)\sqrt{i\nu_l/\omega)}$ contains a large
imaginary part due to the larger imbalance relaxation rate, and the related
mode is highly damped.

In the static limit our equations determine the penetration length $l_E$ of
the electric field into a $d$-wave superconductor in direction $\alpha$, when
a current flows through a contact with a normal metal. Very near T$_c$ when
one can neglect Andreev reflection of the quasiparticles we obtain $l_E=
\sqrt{(\pi\Delta_0 D_\alpha\tau_l)/(4T)},$ which agrees with the results of
Choi \cite{Ch}. Here $D_\alpha$ are diffusion coefficients related to the
conductivities $\sigma_{N\alpha}$ by the relation $D_{\alpha}\kappa^2 =4\pi
\sigma_{N\alpha}$. The anisotropy of $l_E^2$ is proportional to the
conductivity anisotropy.

If the order parameter is not of the pure $d$--type symmetry, but is
close to it: $\langle \Delta(\phi) \rangle^2 \ll \langle
\Delta(\phi)^2 \rangle$, then the results of our calculations are
qualitatively the same. The main distinctions appear in the different energy
and angle dependencies of the quasiparticle relaxation times.

In conclusion, we calculated the linear response of layered $d$-wave
superconductors by means of the kinetic theory. We found the conductivities
determining the quasiparticle currents created by the longitudinal and by the
transverse electromagnetic fields. These results were applied to describe
collective modes and the decay length of the electric field near the boundary
with a normal metal.

We are indebted to U. Eckern for reading the manuscript and helpful comments.


\begin{thebibliography}{20}
\bibitem{Woll93} D. A. Wollman {\em et~al.}, Phys. Rev. Lett. {\bf 71}, 2134 
(1993).
\bibitem{Igu94} I. Iguchi and Z. Wen, Phys. Rev. B {\bf 49}, 12 388 (1994).
\bibitem{Br96} D. A. Browner and H. R. Ott, Phys. Rev. B {\bf 53}, 8249
(1996).
\bibitem{Har93} W. N. Hardi {\em et~al.}, Phys. Rev. Lett. {\bf 70}, 399 
(1993).
\bibitem{Yok} T. Yokoya {\em et~al.}, Phys. Rev. B {\bf 53}, 14 055 (1996).
\bibitem{Sig92} M. Sigrist and T. M. Rice, J. Phys. Soc. Jpn. {\bf 61}, 4283
(1992).
\bibitem{Hir} P. J. Hirschfeld, W. O. Putikka, and D. J. Scalapino, Phys. Rev.
Lett. {\bf 71}, 3705 (1993); Phys. Rev. B {\bf 50}, 10 250 (1996).
\bibitem{TC} M. Tinkham, J.Clarke, Phys. Rev. Lett. {\bf 28}, 1366 (1972).
\bibitem{T} M. Tinkham, Phys. Rev. B {\bf 6}, 1747 (1972).
\bibitem{K} L. V. Keldysh, Zh. Eksp. Teor. Fiz. {\bf 47}, 1515 (1964) [Sov.
Phys. JETP {\bf 20}, 1018 (1977)].
\bibitem{AV} S. N. Artemenko, A. F. Volkov, Uspekhi Fiz. Nauk {\bf 128}, 3
(1979) [Sov. Phys. Usp. {\bf 22}, 295 (1979)]
\bibitem{LO} A. I. Larkin, Yu. N. Ovchinnikov, Zh. Eksp. Teor. Fiz. {\bf 73},
299 (1977) [Sov. Phys. JETP {\bf 46}, 155 (1977)].
\bibitem{A80} S. N. Artemenko, Zh. Eksp. Teor. Fiz. {\bf 79}, 162 (1980) [Sov.
Phys. JETP {\bf 52}, 81 (1980)].
\bibitem{AK} S. N. Artemenko, A. G. Kobel'kov, JETP Letters. {\bf 58}, 445
(1993); Physica C {\bf 253}, 373 (1995)
\bibitem{Gr93} M. Graf, D. Rainer, and J. Sauls, Phys. Rev. B {\bf 47}, 12
089 (1993).
\bibitem{Gr96} M. Graf {\em et~al.}, Phys. Rev. B {\bf 53}, 15 147 (1996).
\bibitem{Ch} C. H. Choi, Phys. Rev. B {\bf 54}, 3044 (1996).
\end{thebibliography}
\end{document}